\documentstyle[tighten,preprint,eqsecnum,aps,prd]{revtex}
\input epsf

\begin{document}
\draft

\def\no{\nonumber}
\def\a{\alpha}
\def\la{\lambda}
\def\be{\begin{equation}}
\def\ee{\end{equation}}
\def\Alpha{{\cal U}}
\def\Beta{{\cal V}}

\preprint{\vbox{\baselineskip=12pt
\rightline{IUCAA-44/97}
\rightline{}
\rightline{gr-qc/9810033}
}}
\title{Behavior of Quasilocal Mass Under Conformal Transformations}
\author{Sukanta Bose\footnote{Electronic address:
{\em sbose@iucaa.ernet.in}}}
\address{Inter-University Centre for Astronomy and Astrophysics, Post Bag 4, 
Ganeshkhind, Pune 411007, India}
\author{Daksh Lohiya\footnote{Electronic address:
{\em dlohiya@iucaa.ernet.in}}}
\address{Inter-University Centre for Astronomy and Astrophysics, 
Post Bag 4, Ganeshkhind, Pune 411007, India \\
and \\ Department of Physics, University of Delhi, Delhi 110007, India}

\maketitle
\begin{abstract}

We show that in a generic scalar-tensor theory of gravity, the ``referenced''
quasilocal mass of a spatially bounded region in a classical solution is 
invariant under conformal transformations of the spacetime metric. We first 
extend the Brown-York quasilocal formalism to such theories to obtain the 
``unreferenced'' quasilocal mass and prove it to be conformally invariant. 
However, this quantity is typically divergent. It is, therefore, essential to 
subtract from it a properly defined reference term to obtain a finite and 
physically meaningful quantity, namely, the referenced quasilocal mass.
The appropriate reference term in this case is defined by generalizing the 
Hawking-Horowitz prescription, which was originally proposed for general 
relativity. For such a choice of reference term, the referenced quasilocal 
mass for a general spacetime solution is obtained. This expression is shown to 
be a conformal invariant provided the conformal factor is a monotonic function 
of the scalar field. We apply this expression to the case of static 
spherically symmetric solutions with arbitrary asymptotics to obtain the 
referenced quasilocal mass of such solutions. Finally, we demonstrate the 
conformal invariance of our quasilocal mass formula by applying it to specific 
cases of four-dimensional charged black hole spacetimes, of both the 
asymptotically flat and non-flat kinds, in conformally related theories.

\end{abstract}

\narrowtext

\vfil
\pagebreak

\section{Introduction}

The lack of a generically meaningful notion of local energy density in 
general relativity is well-known \cite{MTW,ABS,BY}. Essentially, this is 
due to the absence of an unambiguous prescription for decomposing the
spacetime metric into ``background'' and ``dynamical'' components. 
\footnote{See, however, the field formulation of general relativity 
\cite{ff}.} If such a prescription were available, then one could associate 
energy in general relativity with the dynamical component of the metric. In 
the past there have been attempts to define quasilocal energy using
pseudotensor methods \cite{MTW,ABS,LL}. However, these approaches led to 
coordinate-dependent expressions, which lacked an unambiguous geometrical 
interpretation. Another way of defining quasilocal energy has been via
the spinor constructions \cite{Witten,Nester,Penrose,DM}. There are, however,
several unresolved questions regarding this approach, a key issue being the
lack of a rigorous proof of the Witten-Nester integral being a boundary value 
of the gravitational Hamiltonian \cite{Mason}. \footnote{For a more complete 
list of references on quasilocal energy, also see the ones cited in Ref. 
\cite{BY}.} Nevertheless, the total energy of an isolated system has been 
defined in terms of the behavior of the gravitational field at large distances 
from the system \cite{ADM}. Moreover, Brown and York have introduced in Ref. 
\cite{BY} (henceforth referred to as BY) a way to define the quasilocal energy 
of a spatially bounded system in general relativity in terms of the total mean 
curvature of the boundary. Further, in spacetimes with a hypersurface-forming
timelike Killing vector on the boundary of the system, it can be shown that 
there exists a conserved charge, which can be defined to be the quasilocal 
mass associated with the bounded region \cite{BCM}. 

The past few years have seen a revival of interest in scalar-tensor theories
of gravity, primarily, string-inspired four-dimensional dilaton gravity,
which has been shown to yield cosmological as well as charged black hole 
solutions (see Refs. \cite{rev1,rev2,rev3,rev4} for reviews.) 
In particular, the spacetime structure (i.e., geodesics and singularities)
of these black hole solutions, and also those of the Brans-Dicke-Maxwell 
theory (in higher dimensions) \cite{Xan,Cai}, are known to have significant 
differences with respect to the Reissner-Nordstr{\o}m black holes. This 
prompts one to investigate the form of the classical laws of black hole 
mechanics and the ensuing picture of black hole thermodynamics in these 
theories. But the study of the thermodynamical laws entails the knowledge of 
the energy and entropy associated with
these spacetimes. Moreover, {\em equilibrium} thermodynamics of 
a black hole (specifically, in the case of an asymptotically flat solution), 
requires that it be put in a finite-sized ``box'', just as one does in general 
relativity. Thus such a study requires the knowledge of the quasilocal energy
of these ``finite-sized'' systems. 

Recently the BY formalism has been extended to the case of a generic 
scalar-tensor theory of gravity in spacetime dimensions greater than two 
\cite{CM,CCM}. Since solutions of two conformally related scalar-tensor 
theories will themselves be related by a conformal transformation, it 
is interesting to ask if the quasilocal masses of these solutions are also 
related. In the past, it has been suggested that the quasilocal mass is a 
conformal invariant. The reason that is usually provided is that a conformal 
transformation is simply a local field reparametrization, which is supposed 
to leave physical quantities, such as the mass of a system, unchanged. In 
fact, in Ref. \cite{CCM}, Chan, Creighton, and Mann (henceforth referred to 
as CCM) argue that the quasilocal 
mass is indeed invariant under a conformal transformation.

As in general relativity, the (unreferenced) mass of a spacetime in a 
scalar-tensor theory of gravity is typically divergent unless one subtracts a 
(divergent) contribution from a suitable reference geometry to obtain a 
meaningful finite result \cite{BY}. Recently, Hawking and Horowitz (henceforth
referred to as HH) gave a prescription in general relativity for obtaining
an appropriate reference action based on the asymptotic behavior of the fields
on a classical solution \cite{HH}. This reference action is subtracted from 
the original action of the theory to define what is called the physical 
action. The surface term that arises in the Hamiltonian associated with this 
action can be taken as the definition of ``total mass''. This mass turns out 
to be finite and is termed as the physical mass. When evaluated on an 
asymptotically flat spacetime solution, the physical mass of the full 
spacetime coincides with the corresponding ADM expression. 

Following HH, one could ask if a similar prescription can be formulated for
scalar-tensor theories of gravity. If so, it would be interesting to see 
whether the total mass arising from such an action is conformally invariant.
It has been claimed 
by CCM that the generalization of the HH prescription to scalar-tensor 
theories does {\it not} lead to a conformally invariant physical mass. 
They propose their own reference action for a general scalar-tensor theory of
gravity and show that the associated physical mass of static, spherically 
symmetric (SSS) solutions is conformally invariant. However, unlike what
happens in the HH prescription, the CCM reference action is not motivated by
any boundary conditions on the fields that define the spacetime solution of
interest. 

In this paper, we first show that the HH prescription can be generalized to 
the case of scalar-tensor gravity. This reduces the arbitrariness 
in the choice of the reference action. More importantly, we prove that under
certain conditions the 
resulting reference action leads to a conformally invariant referenced 
quasilocal mass. In the following, we will directly deal with only
that conformal transformation that relates the scalar-tensor-gravity metric
to that in the Einstein frame. Of course, our results can be readily
extended to study the behavior of physical quantities under a conformal
transformation relating one scalar-tensor theory to another. In section 
\ref{sec:QE} we derive the expression for the (unreferenced) quasilocal mass 
of a bounded region in $(D+1)$-dimensional spacetime solution of a 
scalar-tensor theory of gravity and prove it to be conformally invariant. 
In section \ref{sec:refac} we generalize the HH prescription 
to the case of scalar-tensor gravity. It is shown that for 
such a choice of the reference action, the referenced quasilocal mass 
is a conformal invariant provided the conformal factor is a monotonic function
of the scalar field. Using this prescription, we give an expression for 
quasilocal mass of static spherically symmetric solutions (with arbitrary 
asymptotics). In section \ref{sec:QEex}, we demonstrate the conformal 
invariance of this quasilocal mass formula by applying it to specific cases of 
four-dimensional (4D) black hole 
spacetimes, of both the asymptotically flat and non-flat kinds, in conformally 
related theories. We briefly summarize and discuss our results in section 
\ref{sec:disc}. In appendix \ref{app:PT}, we show how the standard 
prescription for determining the stress-energy pseudotensor in general 
relativity can be suitably adapted to find the quasilocal energy in 
scalar-tensor theories of gravity. Finally, we demonstrate the consistency of 
the results of the pseudotensor method with the quasilocal formalism of Brown 
and York as applied to scalar-tensor theories.
Throughout this paper, we use the conventions of Misner, Thorne, and Wheeler 
\cite{MTW} and work in ``geometrized units'' $c=1=G$.

\section{Quasilocal Mass under Conformal Transformation}
\label{sec:QE}

Consider a spatially bounded region of a $(D+1)$-dimensional spacetime that is 
a classical solution of a scalar-tensor theory of gravity, such as dilaton 
gravity or Brans-Dicke theory. In this section we extend the formalism of 
Brown and York \cite{BY} to derive an expression for the quasilocal energy of 
gravitational and matter fields associated with such regions. Subsequently, we 
will give an expression for the quasilocal mass.

The BY derivation of the quasilocal energy, as applied to a 
$(D+1)$-dimensional spacetime can be summarized as follows. The system we 
consider is a $D$-dimensional spatial hypersurface $\Sigma$ bounded by a 
$(D-1)$-dimensional spatial hypersurface ${\sf B}$ in a spacetime region that 
can be decomposed as a product of a $D$-dimensional hypersurface and a real 
line interval representing time (see Fig. 1). The time-evolution of the 
boundary ${\sf B}$ is the surface ${}^D {\sf B}$. One can then obtain a 
surface stress-tensor on ${}^D {\sf B}$ by taking the functional derivative of 
the action with respect to the $D$-dimensional metric on ${}^D {\sf B}$. The 
energy surface density is the projection of the surface stress-tensor normal 
to a family of spacelike surfaces like ${\sf B}$ that foliate ${}^D {\sf B}$.
The integral of the energy surface density over such a boundary ${\sf B}$ is 
the quasilocal energy associated with a spacelike hypersurface  $\Sigma$ whose 
{\em orthogonal} ~intersection with ${}^D {\sf B}$ is the boundary ${\sf B}$. 
Here we assume that there are no inner boundaries, such that the spatial 
hypersurfaces $\Sigma$ are complete. In the case where horizons form, one 
simply evolves the spacetime inside as well as outside the horizon.

We follow the same notation as BY. The spacetime metric is $g_{\mu\nu}$ and 
$n^{\mu}$ is the outward pointing unit normal to the surface ${}^D {\sf B}$. 
The metric and extrinsic curvature of ${}^D {\sf B}$ are denoted by 
$\gamma_{\mu\nu}$ and $\Theta_{\mu\nu}$, respectively, and they obey $n^{\mu} 
\gamma_{\mu\nu} =0$ and $n^{\mu} \Theta_{\mu\nu} =0$. Alternatively, 
$\gamma_{\mu\nu}$ and $\Theta_{\mu\nu}$ can be viewed as tensors on ${}^D 
{\sf B}$, denoted by $\gamma_{ij}$ and $\Theta_{ij}$, where $i,j$ refer to 
coordinates in ${}^D {\sf B}$. Similarly, the metric and extrinsic curvature 
of $\Sigma$ are given by the spacetime tensors $h_{\mu\nu}$ and $K_{\mu\nu}$, 
respectively. When viewed as tensors on $\Sigma$, they will be denoted by 
$h_{ij}$ and $K_{ij}$. As in BY, here we will assume that the hypersurface 
foliation $\Sigma$ is ``orthogonal'' to the surface ${}^D {\sf B}$ in the 
sense that on the boundary ${}^D {\sf B}$, the future-pointing unit normal 
$u^{\mu}$ to the hypersurface $\Sigma$ and the outward pointing 
spacelike unit normal $n^{\mu}$ to the surface ${}^D {\sf B}$
satisfy $(u \cdot n) |_{{}^D {\sf B}} =0$. This implies that the shift 
vector, $V^i$, normal to the boundary vanishes, i.e., $V^i n_i =0$. 

\subsection{Action}
\label{subsec:qebd}

We study the following action for a scalar-tensor theory of gravity in 
a $(D+1)$-dimensional spacetime:
\begin{equation}
\label{SBD}
S \left[ \bar{g}_{ab} , \phi, {\cal F}\right] = {1\over 2\kappa}\int d^{(D+1)} 
x \> \sqrt{-\bar{g}} \> U(\phi ) \left[ \bar{R} -  W(\phi) 
(\bar{\nabla} \phi)^2 - V (\phi) +  X(\phi) \bar{L}_{\rm m} 
\right]
\ \ ,
\end{equation}
where $\bar{g}_{ab}$ is the ``physical'' metric, $\phi$ is a scalar field, 
${\cal F}$ represents matter fields, $\kappa \equiv 8\pi$, and
$U$, $V$, $W$, and $X$ are functions of $\phi$. Also, $\bar{L}_{\rm m}$ is the 
matter Lagrangian that includes a possible cosmological constant 
term. The overbar denotes the functional dependence of quantities on the 
physical metric $\bar{g}_{ab}$. Here we assume that $\bar{L}_{\rm m}$ does not 
involve any derivatives of the metric. The dynamics of the 
scalar field is governed by its kinetic term, the effective potential term 
$V(\phi )$ and the non-minimal coupling to the scalar curvature, $\bar{R}$, 
described by $U(\phi )$. The effective potential term can be inclusive of an 
arbitrary additive constant which may occur as a Lagrange multiplier, an 
integration constant, or even a fall out of the renormalization 
procedure of the other matter fields described by $\bar{L}_{\rm m}$. It is 
this constant that is responsible for the ``cosmological constant problem''. 
One can consider the potential term as an effective term obtained by 
integrating over ``heavy'' degrees of freedom as long as one does not go 
beyond the leading semiclassical approximation for the scalar field $\phi$. 

The BY analysis is of course readily applicable in the ``Einstein'' frame, 
which is associated with the ``auxiliary'' metric
\begin{equation}
\label{auxmet}
g_{ab} \equiv U^{2/(D-1)} \bar{g}_{ab}
\ \ ,
\end{equation}
where $D>1$. In terms of the auxiliary metric, one can always cast action 
(\ref{SBD}) as a sum of the Hilbert action, which is independent of the scalar 
and matter fields, and a functional $S_f$ that depends on these fields:
\begin{equation}
\label{SBDaux}
S [g_{ab}] = {1\over 2\kappa}\int d^{(D+1)} x \> \sqrt{-g}\> {R} + S_f
\ \ ,
\end{equation}
where 
\begin{eqnarray}
S_f \equiv {1\over 2\kappa } \int d^{(D+1)} x \>  \sqrt{-g}\> \Bigglb[ &&
{D\over D-1} \left(\nabla \ln U(\phi)\right)^2 - W(\phi) 
(\nabla \phi)^2 \nonumber \\
&&- U^{2/(1-D)} \left( V(\phi) -  X(\phi ) L_{\rm m} \right) \Biggrb]\,.
\label{Sf}
\end{eqnarray}
Above, $L_{m}$ is a functional of the matter fields, their 
derivatives, the auxiliary metric $g_{ab}$, and the scalar field $\phi$. In 
Eq. (\ref{SBDaux}), we have ignored the surface term contributions for the 
present. The subscript $f$ represents the scalar field $\phi$ and the matter 
fields. Note that $S_f$ does not involve any derivatives of the metric. In the
following, $(\bar{g}, \bar{f})$ will denote a field configuration that is a 
solution to (\ref{SBD}), whereas $(g, f)$ is the conformally related solution
in the theory (\ref{SBDaux}). Although there is no bar over $\phi$, note that 
$\phi$ is implicitly included in the configuration $(\bar{g}, \bar{f})$.

\subsection{Quasilocal energy and mass}
\label{subsec:QEaux}

We begin this section by briefly discussing the Hamilton-Jacobi analysis 
used by BY to evaluate the quasilocal energy of a spatially bounded region 
in Einstein gravity. We will later give the expression for the quasilocal 
energy in a generic scalar-tensor theory of gravity.

In general relativity, to make the action functionally differentiable
under boundary conditions that fix the metric on the boundary,
one appends appropriate surface terms to the Hilbert action.
The resulting action in $(D+1)$-dimensions is
\begin{equation}
\label{SBDfunc}
S^1 = {1\over 2\kappa}\int d^{(D+1)} x \> \sqrt{-{g}} \> {R} + {1\over \kappa}
\int_{t'}^{t''} d^D x \>\sqrt{h} \> K - {1\over \kappa} 
\int_{{}^D {\sf B}}
d^D x \>\sqrt{-\gamma} \> \Theta + S_f
\ \ ,
\end{equation}
where $\int_{t'}^{t''} d^D x$ represents the difference of the integral over a 
spatial three-surface $t=t''$ and that over a three-surface $t=t'$. Of course, 
the equations of motion obtained from the variation of the above action are
unaffected by the addition of an arbitrary function $S^0$ of 
fixed boundary data to $S^1$. Hence, the variation of such an action
restricted to classical solutions gives
\begin{eqnarray}
\label{varS}
\delta S_{\rm cl} &=& \{ {\rm terms~involving~variations~in~the~matter~
fields}\} \nonumber \\
&+&  \int_{t'}^{t''} d^D x \> P^{ij}_{\rm cl} \delta h_{ij}
 + \int_{{}^D B} d^D x \> (\pi^{ij}_{\rm cl} - \pi^{ij}_0) \delta \gamma_{ij}
\ \ ,
\end{eqnarray}
where $ P^{ij}$ and $\pi^{ij}$ are, respectively, the momenta conjugate
to $h_{ij}$ and $\gamma_{ij}$, and
\begin{equation}
\label{pi0}
\pi_0 \equiv {\delta S^0 \over \delta \gamma_{ij} } \,.
\end{equation}
Above, the subscript ``${\rm cl}$'' denotes the value of a quantity on a classical 
solution.
As we discuss in detail later, different choices for $S^0$ arise by imposing 
different physical requirements on the quasilocal energy.
 
{}From Eqs. (\ref{varS}) and (\ref{pi0}) given above, we obtain the 
following Hamilton-Jacobi equations:
\begin{eqnarray}
P^{ij}_{\rm cl} |_{t''} &=& {\delta S_{\rm cl} \over \delta h_{ij} (t'')} 
\ \ , \label{P} \\
(\pi^{ij}_{\rm cl} - \pi^{ij}_0 )
&=& {\delta S_{\rm cl} \over \delta \gamma_{ij} } \ .
\label{pi} 
\end{eqnarray}
The quantity that is of interest to us is the surface stress-tensor 
for spacetime and the fields, which is given by
\begin{equation}
\tau^{ij} \equiv {2 \over \sqrt{-\gamma}} {
\delta S_{\rm cl} \over \delta \gamma_{ij}} \ .
\end{equation}
Using (\ref{pi}), we obtain
\begin{equation}
\tau^{ij} = {2 \over \sqrt{-\gamma}}(\pi^{ij}_{\rm cl} - \pi^{ij}_0) \ .
\end{equation}
If $u^i$ is the unit timelike normal to $\Sigma$ on the boundary 
${\sf B}$, then the proper energy surface density $\epsilon$ is
\begin{equation}
\label{energydens}
\epsilon \equiv u_i u_j \tau^{ij} = -{1\over\sqrt{\sigma}} {\delta S_{\rm cl}
\over \delta N} \ \ ,
\end{equation}
where $\sigma_{ij}$ is the metric on the boundary ${\sf B}$. Above, we
made use of the following identity
\be
{\partial \gamma_{ij} \over \partial N} = - {2 u_i u_j \over N} \,.
\ee
Equation (\ref{energydens}) together with (\ref{pi}) can be used to show that 
that the energy surface density is related to the trace of the extrinsic 
curvature, $k_{\rm cl}$, of the boundary ${\sf B}$ embedded in a spatial 
hypersurface $\Sigma$ (which in turn is embedded in a classical solution):
\begin{equation}
\label{energydensaux}
\epsilon = {1\over \kappa} ( k_{\rm cl} - k_0) \ ,
\end{equation}
where $k_0$ is the trace of the extrinsic curvature of a surface that is
{\em isometric} to ${\sf B}$, but is embedded in a reference space. 

Following BY, the extrinsic curvature of ${\sf B}$ as embedded in $\Sigma$ is 
defined by 
\be
\label{kdef}
k_{\mu \nu} = - \sigma^{\alpha}_\mu D_\alpha n_\nu \ \ ,
\ee
where $D_\alpha$ is the covariant derivative on $\Sigma$. Therefore, $k= 
\sigma^{\mu\nu} k_{\mu\nu}$. The quasilocal energy associated with all 
the fields on the spacelike hypersurface $\Sigma$ with boundary 
${\sf B}$ in this ``auxiliary'' spacetime is
\begin{eqnarray}
E &=& \int_{\sf B} d^{(D-1)} x \> \sqrt{\sigma} \> \epsilon = -\int_{\sf B} 
d^{(D-1)} {\delta S_{\rm cl} \over \delta N}\nonumber \\
  &=& {1\over \kappa} \int_{\sf B} d^{(D-1)} x \> \sqrt{\sigma}\>(k_{\rm cl} 
      - k_0) \ .
\label{energyaux}
\end{eqnarray}
In other words, $E$ represents the proper quasilocal energy in the Einstein 
frame. The above expression is interpreted as energy because it is minus the
change in the classical action due to a uniform, unit increase in the proper
time along ${}^D {\sf B}$. Also, for a unit lapse and zero shift, it is equal 
to the Hamiltonian corresponding to the action (\ref{SBDfunc}), as evaluated 
on a classical solution. It is satisfying to note that this is a geometric 
expression independent of the coordinates on the quasilocal surface. However, 
it does depend on the choice of the quasilocal surface and also on the 
foliation of the spacetime by spacelike hypersurfaces. 

When there is a timelike Killing vector field $\xi^\mu$ on the boundary ${}^D 
{\sf B}$, such that it is also hypersurface forming, one can define an 
associated conserved quasilocal mass for the bounded system \cite{BCM,BY}:
\begin{equation}
\label{massaux}
M = \int_{\sf B} d^{(D-1)} x \> \sqrt{\sigma} \> N \epsilon 
\ \ ,
\end{equation}
where $N$ is the lapse function related to $\xi^\mu$ by $\xi^\mu = Nu^\mu$. 
Further, if $\xi \cdot u = -1$, then $N=1$ and 
consequently the quasilocal mass is the same as the quasilocal energy
(\ref{energyaux}). Unlike the quasilocal energy (\ref{energyaux}), the
quasilocal mass is independent of any foliation of the bounded system.

We now ask: What is the analogous expression for the quasilocal energy or 
mass for a bounded spatial region of a spacetime solution of the scalar-tensor 
theory (\ref{SBD})? Note that under boundary conditions 
that fix the metric on the boundary, the appropriate surface action to be
added to the action (\ref{SBD}) is 
\begin{equation}
\label{SBDbdry}
S_{{}^D {\sf B}} [\bar{g}_{ab} ,\phi ] = {1\over \kappa}\int_{t'}^{t''}
 d^{D} x \> \sqrt{- \bar{h}} \> U(\phi) \bar{K}
- {1\over \kappa}\int_{{}^D {\sf B}} 
 d^{D} x \> 
         \sqrt{- \bar{\gamma}} \> U(\phi) \bar{\Theta}
\ \ ,
\end{equation}
where $\bar{K}$ is the trace of the extrinsic curvature of a spatial
hypersurface and $\bar{\Theta}$ that of the boundary ${}^D {\sf B}$ when 
embedded in a spacetime solution of action (\ref{SBD}). It can be verified 
that under a conformal transformation (\ref{auxmet}), the action $S^1$
in (\ref{SBDfunc}) transforms exactly to $S[\bar{g}_{ab}, \phi, {\cal F}] +
S_{{}^D {\sf B}} [\bar{g}_{ab} ,\phi ]$ , which is defined by Eqs. (\ref{SBD})
and (\ref{SBDbdry}). As in the case of 
Einstein gravity discussed above, one can use the
BY approach to derive the expression for quasilocal energy from the above
surface action in a non-minimally coupled theory. Such a calculation was
done by Creighton and Mann \cite{CM} for four-dimensional pure 
dilaton-gravity. A straightforward generalization of their derivation to 
the case of a $(D+1)$-dimensional scalar-tensor theory (\ref{SBD}) including 
matter fields gives the quasilocal energy in such theories to be
\begin{eqnarray}
\bar{E} &=& \int_{\sf B} d^{(D-1)} x \> \sqrt{\bar{\sigma}} \>
            \bar{\epsilon } \nonumber \\
&=& {1\over \kappa} \int_{\sf B} d^{(D-1)} x \> \sqrt{\bar{\sigma}} \> 
 \left( U(\phi ) \bar{k} - \bar{n}^i \partial_i U(\phi ) \right) 
    -\bar{E}_0  \,.
\label{energyBD}
\end{eqnarray}
In the next section we will consider appropriate reference actions $S^0$ and 
their respective contributions, $\bar{E}_0$, to the above expression. In the 
appendix we give an alternative derivation of the above
energy expression using the pseudotensor method.

Analogous to Eq. (\ref{massaux}) one can also define the quasilocal mass
in the scalar-tensor theory to be
\begin{eqnarray}
\label{massBD}
\bar{M} &=& \int_{\sf B} d^{(D-1)} x \> \sqrt{\bar{\sigma}} \>
                 \bar{N} \bar{\epsilon} \no \\
&=& {1\over \kappa} \int_{\sf B} d^{(D-1)} x \> \sqrt{\bar{\sigma}} \> 
 \bar{N} \left( U(\phi ) \bar{k} - \bar{n}^i \partial_i U(\phi ) \right) 
    -\bar{M}_0
\ \ ,
\end{eqnarray}
where $\bar{M}_0$, is an appropriate reference term.

\subsection{Conformal transformation}
\label{subsec:conftrans}

We now study how the quasilocal mass, ${\bar M}$, modulo the reference term 
${\bar M}_0$, behaves under a conformal transformation. Equation 
(\ref{massaux}) shows that this requires a knowledge of how the total mean 
curvature $k$ of the boundary ${\sf B}$ behaves under a conformal 
transformation. Let the physical metric $\bar{g}_{ab}$ of the 
scalar-tensor theory be related to the auxiliary metric $g_{ab}$ by the 
conformal transformation
\begin{equation}
\label{confmet}
\bar{g}_{ab} \equiv \Omega^2 g_{ab} \ \ ,
\end{equation}
where $\Omega$ is generally a function of the spacetime coordinates. 
Comparing with (\ref{auxmet}), we find that 
\be \label{OU}
\Omega = U^{-1/(D-1)} \,.
\ee 
Note that on-shell $U(\phi)$ will be determined through the equations of 
motion pertaining to the action (\ref{SBD}).

Let us embed the $(D-1)$-dimensional spatial boundary ${\sf B}$ in each of the 
two conformally related spacetimes, assuming that such embeddings are feasible 
and unique. Then the unit timelike normal $\bar{u}^i$ in the physical
spacetime is related to that in the auxiliary spacetime, $u^i$, as follows
\begin{equation}
\label{normal1}
\bar{u}^i = \Omega^{-1} u^i
\ \ .
\end{equation}
Similarly, the outward pointing unit normal to the surface
${}^D {\sf B}$ in the two spacetimes are related as 
\begin{equation}
\label{normal2}
\bar{n}^i = \Omega^{-1} n^i
\ \ .
\end{equation}
One can further show that the extrinsic and the total mean curvature of the 
boundary ${\sf B}$, as embedded in these spacetimes, are related as follows:
\begin{eqnarray}
\bar{k}_{ij} &=& \Omega \left[ k_{ij} - \sigma_{ij} n^l \nabla_l (\ln \Omega)
                      \right] \ ,
\label{barkij} \\
\bar{k} &=& \Omega^{-1} \left[ k - (D-1) n^l \nabla_l (\ln \Omega)
                      \right] \ ,
\label{bark}
\end{eqnarray}
where $n^l$ is the spacelike unit normal to the surface ${}^D {\sf B}$ 
(embedded in the auxiliary spacetime). Formally, we associate the covariant
derivatives $\nabla_l$ and $\bar{\nabla}_l$ with metrics $g_{ab}$ and $\bar{
g}_{ab}$, respectively. In spacetime regions where $\Omega$ is non-singular, 
one can invert the above relation to obtain
\begin{equation}
\label{k}
k = \Omega {\bar k} + (D-1) \Omega \>\bar{n}^l \bar{\nabla}_l 
       (\ln \Omega) \ \ ,
\end{equation}
where $k$ is the total mean curvature of the boundary ${\sf B}$, as
embedded in the auxiliary spacetime.

Equation (\ref{k}) shows that under a conformal transformation the quasilocal
mass defined in Eq. (\ref{massaux}), modulo the reference term arising from 
$k_0$, transforms as follows:
\be
{1\over \kappa} \int_{\sf B} d^{(D-1)} x \> N \sqrt{\sigma} \> k 
= {1\over \kappa}\int_{\sf B} d^{(D-1)} x \>\bar{N} 
       \sqrt{\bar{\sigma}} \left[ U(\phi) \bar{k} - \bar{n}^i 
       \partial_i U(\phi) \right] \ \ ,
\label{CTbdryac}
\ee
where we have used Eq. (\ref{OU}). Applying the above identity to the mass
expressions (\ref{massaux}) and (\ref{massBD}) proves that the quasilocal 
masses of conformally related spacetimes are the same, provided the reference 
term $\bar{M}_0$ is conformally invariant. 

Consider the behavior of the timelike vector $\xi^\mu$ defined above Eq.
(\ref{massaux}). It is assumed to be Killing in a given frame, say, the 
Einstein 
frame. It is also a conformal invariant, i.e., $\bar{\xi}^\mu = \bar{N}\bar{
u}^\mu = Nu^\mu = \xi^\mu$. However, it will not remain Killing in a general 
conformal transformation. Thus, although the (unreferenced) quasilocal mass is 
a conformal invariant, its property of being a conserved charge in a given 
frame is not. However, if $\xi^\mu$ obeys $\xi^\mu \nabla_\mu \phi =0$, then it
continues to remain Killing in the conformally related frame. In such an
event, the associated quasilocal mass remains a conserved charge in that frame
too. Furthermore, the unreferenced quasilocal energy is not invariant under
a conformal transformation. When $\Omega$ is independent of the coordinates on 
the quasilocal surface, it transforms as 
\be \label{qEct}
\bar{E} -\bar{E}_0 = \Omega^{-1} (E-E_0)
\ee
and, hence, is not a conformal invariant.

Finally, note that when we compare the quasilocal masses of conformally 
related spacetimes above, we assume that the boundary ${}^D {\sf B}$,
which is taken to be embedded in a particular spacetime, is also
embeddable in the conformally related spacetime. However, the embeddability 
of a hypersurface requires 
that the intrinsic and extrinsic geometry of the boundary obey the 
Gauss-Codazzi, Codazzi-Mainardi, and Ricci integrability conditions in both 
spacetimes separately. In general, not all of these integrability conditions 
are conformal invariants. Therefore, embeddability 
of a hypersurface in a spacetime does not guarantee its embeddability in a
conformally related spacetime. Nevertheless, it can be shown that one can 
always embed a $(D-1)$-dimensional spacelike spherical boundary in 
$(D+1)$-dimensional SSS spacetime solutions, which are Ricci flat, and
in spacetimes related through conformal transformations that preserve these
spacetime properties \cite{CCG,TJW}.

\section{Reference action and quasilocal mass}
\label{sec:refac}

The Brown-York definition of the quasilocal energy (\ref{energyaux}) 
associated with a spatially bounded region of a given spacetime solution is 
not unique. This is because an arbitrary functional $S^0$ of the boundary data 
can be added to the action without affecting the equations of motion. On the 
other hand, to get a well-defined (finite) expression for the quasilocal 
energy of spatially non-compact geometries, one is usually required to 
subtract the (divergent) contribution of some reference background. 
At the level of the action such a ``regularization'' is tantamount to the 
addition of a reference action $S^0$, which is a functional of appropriate 
background fields $(g_0 , f_0 )$, to the original action $S^1$. 
For 4D Einstein gravity, BY prescribe the following reference action
\begin{equation}
\label{BYref}
S^0 = -\int_{ {}^3 {\sf B}} d^3 x \> \left[ N \sqrt{\sigma } (k/ \kappa )|_0
    +  2\sqrt{\sigma} V^a (\sigma_{ai} n_j P^{ij} / \sqrt{h} )|_0 \right]
\ \ ,
\end{equation}
which is a linear functional of the lapse $N$ and shift $V^a$. Above,
${}^3 {\sf B}$ is the time-evolution of a two-boundary ${\sf B}$ that
is embedded in a fixed three-dimensional spacelike slice $\Sigma$ of some 
fixed reference spacetime. Also, $k|_0$ and $(\sigma_{ai} n_j P^{ij} / 
\sqrt{h} )|_0$ are arbitrary functions of the two-metric $\sigma_{ab}$ on 
the boundary ${\sf B}$, $n^j$ is the unit normal to the 2-boundary ${\sf B}$,
and $\{ h_{ij} , P^{ij} \}$ are the canonical 3-metric and the conjugate
momentum on the three-dimensional spacelike slice $\Sigma$. Varying the
lapse in the first term in (\ref{BYref}) gives the energy surface density,
whereas varying the shift in the second term gives the momentum surface 
density in the reference spacetime \cite{BY}. Since we mainly discuss the 
application of (\ref{BYref}) to evaluate the {\em proper} quasilocal mass or 
energy, which is obtained by the variation of the total action (on classical 
solutions) with respect to $N$, we will henceforth drop the last term in 
(\ref{BYref}) from our consideration. 

To calculate the quasilocal energy associated with regions of spacetime 
solutions of $(D+1)$ -dimensional Einstein gravity, the appropriate 
generalization of the BY reference action is again given by (\ref{BYref}),
except that the integration is now over the boundary ${}^D {\sf B}$. The 
boundary ${}^D {\sf B}$ itself is the time-evolution of the 
$(D-1)$-dimensional spatial boundary ${\sf B}$. For asymptotically flat 
spacetimes an appropriate reference background might be
vacuum flat spacetime. 

However, for an arbitrary spacetime solution (eg., spacetimes that are neither
spatially closed nor asymptotically flat), a more well defined prescription 
for the choice of $S^0$ is required. Recently, one such prescription was 
given by Hawking and Horowitz in their quest for obtaining the total mass of 
spacetimes with arbitrary asymptotic behavior in general relativity 
\cite{HH}. Their starting point is the ``physical'' action defined as
\begin{equation}
\label{HHref}
S_P (g, f) \equiv S (g, f) - S (g_0, f_0)
\ \ ,
\end{equation}
where $(g_0, f_0)$ are fields specifying a reference static background, which
is a {\em solution} to the field equations. Therefore, the physical action
of the reference background is zero. Given a solution $(g,f)$, in order to 
determine a reference background, $(g_0, f_0)$, HH fix a three-boundary 
(${}^3 {\sf B}$) near infinity and require that $(g,f)$ induce the same fields 
on this boundary as $(g_0, f_0)$. The energy of a solution can be obtained 
from the physical Hamiltonian associated with $S_P$ (for details, see Ref. 
\cite{HH}) and is similar to the BY quasilocal expression. For asymptotically 
flat spacetime solutions, the reference background is chosen to be flat 
space and the resulting energy expression agrees with the 
one obtained in the ADM formalism.

It is important to note that the HH prescription allows one to compute the
total energy associated with a general time translation $t^\mu = Nu^\mu +
V^\mu$. In a generic case, the resulting energy will have a shift-dependent
contribution, such as the second term in (\ref{BYref}). However, such a term
vanishes when the spacetime is taken to approach a static 
background solution and the resulting expression (with $N=1$) is the same as 
the BY energy (\ref{energyaux}). Even if the spacetime is asymptotically 
non-static this term will vanish when $V^a \sigma_{ab} =0$. This happens, eg.,
for cosmological solutions with the Robertson-Walker metric.

Building on the work of Brown and York, Chan, Creighton, and Mann 
\cite{CCM,KCK} chose a particular reference action to compute the
quasilocal masses of solutions in scalar-tensor theories. In the special case 
of SSS spacetimes, it has been shown by CCM that their
choice leads to a conformally invariant referenced quasilocal mass. 
A second possibility of obtaining a reference action is to generalize the HH 
prescription to scalar-tensor theories. Such an attempt was also made by CCM 
\cite{CCM}. However, they conclude that the mass formula obtained using their 
generalization of the HH prescription is not conformally invariant. For 
details on this issue, we refer the reader to Ref. \cite{CCM}.

In this section, we extend the BY formalism to obtain the referenced 
quasilocal mass associated with bounded regions of spacetime solutions (with 
arbitrary asymptotic behavior) in scalar-tensor gravity. A relevant question 
in such an analysis is whether the reference action can be specified in a 
unique way. Finding an answer to this would in itself be an interesting pursuit
and involves addressing issues of positivity of the mass or energy of such 
solutions as well as the stability of the corresponding reference solution. 
Here, we do not attempt to find if the reference action or solution can be 
uniquely specified at all. Below, after discussing the CCM analysis briefly, 
we present our alternative generalization of the HH prescription to 
scalar-tensor gravity. Although, this does not select a unique reference 
action, nevertheless, invoking this prescription reduces the number of 
allowed reference actions. We prove that under certain conditions such a 
prescription does lead to a conformally invariant referenced quasilocal mass. 

\subsection{The CCM prescription}
\label{subsec:CCMpres}

For a non-minimally coupled action of the type (\ref{SBD}), the reference
action suggested by CCM is \cite{CCM}
\begin{equation}
\label{CCMref}
S^0 = -\int_{{}^D {\sf B}} d^D x \> \bar{N} \sqrt{\bar{\sigma}} U(\phi) 
(\bar{k}_{\rm flat} / \kappa )   
\ \ ,
\end{equation}
where $\bar{k}_{\rm flat}$ is the trace of the extrinsic curvature of the 
$(D-1)$-boundary ${\sf B}$ embedded in a $D$-dimensional flat spatial slice.
Consider the special case of an asymptotically flat SSS 
spacetime metric, as a solution in this theory: 
\begin{equation}
\label{SSS}
d\bar{s}^2 = - \bar{N}^2 (r) dt^2 + {dr^2 \over \bar{\la}^2 (r)} + 
                  r^2 d \omega^2
\ \ ,
\end{equation}
where $\bar{N}$ and $\bar{\la}$ are functions of $r$ only, and $d \omega^2$ is 
the line element on a unit $(D-1)$-sphere. Let us now make the following 
conformal transformation
\begin{equation}
\label{CT1}
\tilde{g}_{ab} = \tilde{\Omega}^2 \bar{g}_{ab} \quad , \quad 
\tilde{U} = \tilde{\Omega}^{(1-D)} U \ \ ,
\end{equation}
where $U(\phi)$ is the scalar-field dependent coupling appearing in
(\ref{CCMref}). Note that under this conformal transformation the functional 
form of $S^0$,
\be \label{CCMref2}
S^0 = -\int_{{}^D {\sf B}} d^D x\>\tilde{N} \sqrt{\tilde{\sigma}} 
\tilde{U}(\phi) (\tilde{k}_{\rm flat} / \kappa )   
\ \ ,
\end{equation}
remains unchanged provided we assume $\tilde{N} = \tilde{\Omega} \bar{N}$. 

Let the metric (\ref{SSS}) be related through this 
conformal transformation to the following SSS metric:
\begin{equation}
\label{SSSc}
d\tilde{s}^2 = - \tilde{N}^2 (r) dt^2 + {dr^2 \over \tilde{\la}^2 (r)} + 
                 \tilde{\Omega}^2 r^2 d \omega^2 \ \ ,
\end{equation}
which is assumed to arise as a solution to another scalar-tensor theory that 
is related to action (\ref{SBD}) by the conformal transformation (\ref{CT1}).
Above, $\tilde{N} = \tilde{\Omega} \bar{N}$ and $\tilde{\la}=\tilde\Omega^{-1} 
\bar{\la}$, where $\tilde{\Omega}$ is a function of $r$ only. 

For the special case of the spacetime solution (\ref{SSS}), and with the 
choice of reference action (\ref{CCMref}), CCM argue that the quasilocal mass 
associated with the region inside a sphere of curvature radius $r$, which is 
embedded in spacetime (\ref{SSS}), can be expressed as
\begin{equation}
\label{CCMM}
\bar{M}(r) = {\bar{N}(r) \over \kappa} \left( { (D-1) \bar{A}_{D-1} (r) 
U(\phi) \over  r} - \bar{\la}(r) {d\over dr} \left( \bar{A}_{D-1} 
(r) U(\phi)  \right) \right) \,.
\end{equation}
Above, $\bar{A}_{D-1}$ is the area of the boundary $(D-1)$-sphere of radius 
$r$ given by
\begin{eqnarray}
\bar{A}_{n} &=& \int_{\sf B} d^{n} x \> \sqrt{\bar{\sigma}} \no\\
     &=& {(4\pi)^{n/2} \Gamma (n/2) \over \Gamma (n)} r^n \ \ , \label{Sarea}
\end{eqnarray}
$\bar{\sigma}_{ij}$ being the metric on ${\sf B}$. Note that the first term in 
Eq. (\ref{CCMM}) is just the reference term 
\begin{equation}
\label{CCMrefterm}
\bar{M}^0 = -\int_{{}^D {\sf B}} d^{D-1} x \> \bar{N} \sqrt{\bar{\sigma}} 
U(\phi) (\bar{k}_{\rm flat} / \kappa )   
\ \ ,
\end{equation}
whereas the second term arises from the quasilocal mass definition 
(\ref{massBD}) on using the identity
\begin{equation}
\label{kA}
{1\over \kappa} \int_{\sf B} d^{(D-1)} x \> \sqrt{\bar{\sigma}} \> 
 \bar{N}  \bar{k} = -{\bar{N} (r) \bar{\la} (r) \over \kappa} {d\over dr}
\bar{A}_{(D-1)} (r) \ \ ,
\end{equation}
which holds for the SSS metric (\ref{SSS}). We will call Eq. (\ref{CCMM}) the 
CCM mass expression. Similarly, CCM find that for the metric (\ref{SSSc}), the
quasilocal mass is
\begin{equation}
\label{CCMMc}
\tilde{M}(r) = {\tilde{N}(r) \over \kappa} \left( { (D-1) \tilde{A}_{D-1} (r) 
\tilde{U} \over \tilde{\Omega} r} - \tilde{\la}(r) {d\over dr} \left( 
\tilde{A}_{D-1} (r) \tilde{U}  \right) \right) \ \ ,
\end{equation}  
where $\tilde{A}_{(D-1)} = \tilde{\Omega}^{(D-1)} \bar{A}_{(D-1)}$. 

Thus, the CCM mass $\bar{M}(r)$ defined in Eq. 
(\ref{CCMM}) is invariant under the conformal transformation (\ref{CT1}), 
namely, $\bar{M}(r) = \tilde{M}(r)$. To be precise, each term 
in (\ref{CCMM}) is separately conformally invariant. Finally, let us emphasize
that unlike in the HH prescription, in the CCM prescription one does not
require the `background' fields appearing in the reference action 
(\ref{CCMref}) to constitute a solution of that action. Also, the choice of
the CCM reference action is independent of the asymptotic behavior of the
fields of the solution. (This is the reason why the referenced quasilocal mass 
(with $N=1$) in this prescription differs from the Abbott-Deser definition 
of the total energy \cite{AD} when applied to asymptotically anti-de Sitter 
SSS spacetimes. One can, however, recover this energy expression by 
generalizing the CCM reference action to the case of such spacetimes (for 
details, see Ref. \cite{BCM}).)

\subsection{An alternative prescription for reference action
and quasilocal mass}
\label{subsec:BL}

First, we extend the applicability of the HH prescription to scalar-tensor 
gravity in order to obtain an appropriate reference action. Given a solution, 
$(\bar{g},\bar{f})$, one chooses a reference background {\em solution}, $(\bar{
g}_0 ,\bar{f}_0 )$, by using the HH prescription as enunciated above in this 
section. Then the appropriate reference action is simply
\be \label{HHrefac}
\left[ S[\bar{g}_{ab}, \phi , {\cal F}]+S_{{}^D {\sf B}} [\bar{g}_{ab}, \phi]
\right]_{\rm ref} \ \ ,
\ee
where $S$ and $S_{{}^D {\sf B}}$ are given by Eqs. (\ref{SBD}) and 
(\ref{SBDbdry}), respectively, and $[{\rm term}]_{\rm ref}$ denotes the value 
of the term as evaluated on the reference solution. In 
general, the reference action can depend on the initial and final metrics 
$\bar{h}_{ij} (t')$ and $\bar{h}_{ij} (t'')$ through spatial boundary terms, 
namely, the first term on the right-hand side of Eq. (\ref{SBDbdry}). 
However, in the present calculation such contributions can be dropped since 
they do not affect the BY quasilocal mass.

Second, we address the question: If the HH prescription is obeyed by a pair 
of solutions, $(\bar{g},\bar{f})$ and $(\bar{g}_0 ,\bar{f}_0 )$, for the 
boundary ${\sf B}$ in a given frame, then, will it also be obeyed by  
conformally related fields in a conformally related frame? We answer this as 
follows. Note that the reference solution $(\bar{g}_0
, \bar{f}_0 )$ is conformally related to that in Einstein gravity, $(g_0 , f_0 
)$, by Eq. (\ref{confmet}), where $U$ is now a function of $\phi_0$. Here, 
both $U(\phi_0 )$ and the conformal factor $\Omega$ are positive-definite
quantities. Thus, for a solution in scalar-tensor gravity, $(\bar{g},\bar{f})$,
if the lapse $\bar{N}$ and the fields $(\bar{\sigma}_{ab} , \phi)$ induced on 
the boundary ${\sf B}$ match with the lapse $\bar{N}_0$ and the fields 
$(\bar{\sigma}_{0ab} , \phi_0 )$ at ${\sf B}$ in the reference spacetime, then 
for the conformally related configuration $(g, f)$ in the Einstein frame, the 
lapse $N$ and the field $\sigma_{ab}$ at ${\sf B}$ will necessarily match with 
their reference spacetime counterparts $N_0$ and $\sigma_{0ab}$ induced on the 
corresponding boundary. This holds provided $\Omega$ is a monotonic function of
$\phi$. To repeat, let 
\be \label{proofBD}
\bar{N} |_{\sf B} = \bar{N}_0 |_{\sf B} \>, \quad 
\bar{\sigma}_{ab} |_{\sf B} = \bar{\sigma}_{0ab} |_{\sf B}, \quad
\phi |_{\sf B} = \phi_0 |_{\sf B} \,.
\ee
Then, using the above conditions, we can infer the following requirements on
the Einstein frame fields:
\begin{eqnarray} \label{proofEF}
N|_{\sf B} &=& \left[\bar{N} \Omega^{-1} (\phi)\right]_{\sf B} = 
\left[ \bar{N}_0 \Omega^{-1} (\phi_0 ) \right]_{\sf B} = N_0 |_{\sf B} \ \ ,
\nonumber \\
\sigma_{ab} |_{\sf B} &=&\left[ \bar{\sigma}_{ab} \Omega^{-2} (\phi)
\right]_{\sf B} = \left[\bar{\sigma}_{0ab} \Omega^{-2} (\phi_0 ) 
\right]_{\sf B} = \sigma_{0ab} |_{\sf B} \,.
\end{eqnarray}
This proves that, for such a conformal factor, if the HH 
prescription is obeyed in a given frame, say, the scalar-tensor frame, it will 
automatically be satisfied in the Einstein frame. It is easy to extend this
proof to the case of any two conformally related frames.

A meaningful referenced quasilocal mass can now be defined. It is simply given 
by Eq. (\ref{massBD}), where the reference term, $\bar{M}_0$, is obtained 
from the HH prescribed reference action (\ref{HHrefac}) by a BY type 
analysis (as described in section \ref{subsec:QEaux}), i.e.,
\be
\label{massBDref}
\bar{M}_0= {1\over \kappa} \int_{\sf B} d^{(D-1)} x \> \sqrt{\bar{\sigma}_0} 
\>\bar{N}_0 \left( U(\phi_0 ) \bar{k}_0 - \bar{n}^i_0 \partial_i U(\phi_0 ) 
\right) 
\ \ ,
\ee
which is just the first term on the right-hand side of Eq. (\ref{massBD}) as
evaluated on the reference solution.

We now show that the referenced quasilocal mass so obtained is 
conformally invariant. In the previous section, we proved that 
the unreferenced quasilocal mass is a conformal invariant. What remains to be 
verified is that the reference term $\bar{M}_0$ is also invariant under the 
transformation $\bar{g}_{0ab} = \Omega (\phi_0) g_{0ab}$. This is easily done
by applying the curvature-transformation identity (\ref{CTbdryac}) to the 
above expression for $\bar{M}_0$. This shows that $\bar{M}_0 = M_0$, 
which proves the conformal invariance of the reference term. Hence the 
referenced quasilocal mass is conformally invariant. 
(This, of course, presumes a monotonic $\Omega$.)

We now illustrate this invariance explicitly for the case of SSS spacetimes.
By applying the mass expressions (\ref{massBD}) and (\ref{massBDref})
to the SSS metric (\ref{SSS}), we obtain
\begin{equation}
\label{BLM}
\bar{M}(r) =  \left[ {\bar{N}(r) \over \kappa} \bar{\la}(r)
    {d\over dr} \left( \bar{A}_{D-1} (r) U(\phi) \right) \right]^0_{\rm cl}
\ \ ,
\end{equation}
where $\bar{A}$ is given in Eq. (\ref{Sarea}) and $[{\rm term}]^0_{\rm cl}$ 
is defined as the difference in the values of the term evaluated on the 
reference spacetime and on the spacetime solution whose mass we aim to 
compute. Note that in keeping with the HH prescription, we require that at the 
boundary, $r=r_B$, the SSS solution satisfies $\bar{N}(r_B) = \bar{N}_0 
(r_B)$, $\bar{\sigma}_{ab}(r_B) = \bar{\sigma}_{ab}(r_B)|_0$, and $U(\phi(r_B)
) = U(\phi_0 (r_B))$. To obtain the total mass of an asymptotically flat 
spacetime, one first evaluates $\bar{M}(r)$ for general $r$ and then imposes 
the limit $r\rightarrow \infty$. In this limit, Eq. (\ref{BLM}) yields the ADM 
mass when the reference solution is chosen to be flat. The referenced 
quasilocal mass defined in Eq. (\ref{BLM}) is manifestly invariant under the 
conformal transformation (\ref{CT1}) and, therefore, is the same as the 
expression obtained upon removing the overbars in that equation. 

Alternatively, consider applying the mass expressions (\ref{massBD}) and 
(\ref{massBDref}) to an SSS metric of the form (\ref{SSSc}), namely,
\begin{equation}
\label{SSSc2}
d\bar{s}^2 = - \bar{N}^2 (r) dt^2 + {dr^2 \over \bar{\la}^2 (r)} + 
                 {\Omega}^2 r^2 d \omega^2 \,.
\end{equation}
It is easy to verify that the resulting quasilocal mass expression is 
identical to Eq. (\ref{BLM}). However, the area of ${\sf B}$ (as embedded in
metrics of the type (\ref{SSSc2})) is now given as
\begin{eqnarray}
\bar{A}_{n} &=& \int_{\sf B} d^{n} x \> \sqrt{\bar{\sigma}} \no\\
     &=& {(4\pi)^{n/2} \Gamma (n/2) \over \Gamma (n)} r^n {\Omega}^n
\,. \label{Sareac}
\end{eqnarray}
Here too the referenced quasilocal mass (\ref{BLM}) remains invariant under 
conformal transformations of the metric (\ref{SSSc2}), provided the HH
prescription is followed in determining the reference solution.

To summarise, we define the referenced quasilocal mass of a solution 
associated with a boundary ${\sf B}$ as the difference of its unreferenced 
quasilocal mass from that of a reference field configuration, which is also a 
solution of the theory and obeys the HH prescription. Under a conformal
transformation, this pair of solutions has its ``image'' pair, which comprises
of two solutions in the conformally related frame; in that frame, the 
referenced quasilocal mass is again the difference of the unreferenced 
quasilocal masses of these two image solutions. To investigate the behavior of 
the referenced quasilocal mass under a conformal transformation, one must
therefore study how the unreferenced quasilocal masses of these two solutions
transform under the conformal map ${\bar g}_{ab} = \Omega(\phi) g_{ab}$ and 
${\bar g}_{0ab} = \Omega(\phi_0) g_{0ab}$, {\em respectively}. Such a study
reveals the conformal invariance of our referenced quasilocal mass (\ref{BLM}).

We end this section by noting that, when applied to the case of asymptotically
flat SSS spacetimes, there is a subtle but significant 
difference between the quasilocal mass definition (\ref{BLM}), which we 
propose above, and the mass definition that CCM obtain by their 
generalization of the HH prescription \cite{CCM}, namely,
\be
\label{HHCCM}
\bar{M}(r) =  \left[ {\bar{N}(r) \over \kappa} \left( 1-\bar{\la}(r) \right)
         {d\over dr} \left( \bar{A}_{D-1} (r) U(\phi) \right) \right]_{\rm cl}
\,.
\end{equation}
Specifically, consider the case of SSS metrics of the form (\ref{SSS}). Then,
the above formula can be obtained from (\ref{BLM}) in two steps. First, one 
sets $\bar{\la}(r)|_0 =1$ in Eq. (\ref{BLM}). This can always be done, for, 
the reference spacetime solution in such a case is flat. Second, and more 
importantly, one {\em assumes} that
\be
\label{HHCCMcond}
{d U(\phi_{\rm cl}) \over dr} = {d U(\phi_0) \over dr}
\ \ , \ee
at the boundary ${\sf B}$. This, however, is an additional requirement over and
above those included in the HH prescription. Consequently, Eq. (\ref{HHCCM}) 
is different from Eq. (\ref{BLM}), where condition (\ref{HHCCMcond}) is not 
assumed. This is also the reason why Eq. (\ref{HHCCM}), as opposed to Eq. 
(\ref{BLM}), fails to be conformally invariant.

In the next section we apply our mass definition (\ref{BLM}) to find 
the referenced quasilocal masses of charged black holes in 4D dilaton gravity, 
and their conformally related cousins in 4D Einstein gravity.

\section{Quasilocal mass in scalar-tensor theories of gravity: Examples}
\label{sec:QEex}

\subsection{Asymptotically flat SSS spacetimes}
\label{subsec:4Ddilgrav}

Let us consider the charged black hole solutions of the four-dimensional 
dilaton gravity action (see Refs. \cite{rev1,rev2} for reviews)
\begin{equation}
\label{S4Ddil}
S = {1\over 2\kappa} \int d^4 x \> \sqrt{-\bar{g}}e^{-2\phi}
  [ \bar{R}+ 4(\bar{\nabla} \phi)^2 - 2\Lambda - \bar{F}^2] 
\ \ ,
\end{equation}
where $\bar{R}$ is the four-dimensional Ricci scalar, $\Lambda$ is a 
cosmological constant and $\bar{F}_{\mu\nu}$ is the Maxwell field associated 
with a U(1) subgroup of ${\rm E}_8 \times {\rm E}_8$. In this subsection we 
will consider the case where $\Lambda = 0$. The magnetically charged black 
hole solution to the above action is \cite{GM,GHS}
\begin{eqnarray}
d \bar{s}^2 &=& -{e^{2\phi_\infty}(1-2m e^{\phi_\infty} /r) 
\over (1-Q^2  e^{-\phi_\infty} /(mr))}dt^2 \nonumber \\    
&&+ {dr^2 \over (1-2me^{\phi_\infty} /r) (1-Q^2 e^{-\phi_\infty} /(mr))}
 +r^2 d\omega^2 \ \ ,
\label{4Ddilbh} \\
e^{-2\phi} &=& e^{-2\phi_\infty} \left( 1 - {Q^2 e^{-\phi_\infty} \over mr}
 \right) = U(\phi)\ \ ,
\label{4Ddil} \\
\bar{F}&=&Q\sin \theta d\theta \wedge d\phi \ \ ,
\label{4DF}
\end{eqnarray}
where $m$ and $Q$ are classical hairs of the stringy black hole and 
$\phi_\infty$ is the asymptotic constant value of the dilaton. Above, $m$ is 
also called the Schwarzschild mass of the spacetime and $Q$ is the magnetic 
charge of the black hole. The strings couple to the above metric, 
$\bar{g}_{\mu\nu}$, as opposed to the one related through the conformal 
transformation $g_{\mu\nu} \equiv e^{-2\phi} \bar{g}_{\mu\nu}$, which casts 
the above action in the Hilbert form. 

We will now demonstrate that the quasilocal mass of a spatial region enclosed
inside the two-sphere of curvature radius $r_B$ is conformally invariant. We
first calculate the mass in the string frame. Since the spacetime 
(\ref{4Ddilbh}) is asymptotically flat, we choose the reference metric to be 
flat\footnote{see the discussion in section \ref{sec:disc}}:
\be
\label{GHS-ref}
d \bar{s}^2_0 = -\bar{N}^2_0 {dt}^2  + dr^2 +{r}^2 d\omega^2
\ \ , \ee
where $\bar{N}_0$ is a constant. Note that the above metric is a solution of
the action (\ref{S4Ddil}) with $\phi_0 =$ constant and $\bar{F}_0 = 0$. A 
two-sphere boundary of curvature radius $r=r_B$ can be isometrically
embedded in both the above spacetimes (\ref{4Ddilbh}) and (\ref{GHS-ref}).
For the lapse at the boundary to match in these spacetimes, we choose
\be
\bar{N}_0 = {e}^{\phi_\infty} \left(1- {2me^{\phi_\infty} \over r_B}
\right)^{1/2}  \left(1- {Q^2 e^{-\phi_\infty} \over mr_B} \right)^{-1/2} \,.
\ee
For the remaining HH requirement to be satisfied, the value of $\phi$ induced
at the boundary in these spacetimes should match. This implies that on the
reference spacetime (\ref{GHS-ref}), one must have 
\be
\label{phi-string}
e^{-2\phi_0}  = e^{-2\phi_\infty} \left(1- {Q^2 e^{-\phi_\infty} \over 
mr_B}\right) = U(\phi_0 ) 
\ee
everywhere. Using these expressions in (\ref{BLM}), we find that the 
quasilocal mass is
\begin{eqnarray}
\bar{M}(r_B) &=& e^{-\phi_\infty} r_B \Bigg[
-\left(1- {Q^2 e^{-\phi_\infty} \over mr_B}\right) \left(1-{2me^{\phi_\infty} 
\over r_B}\right)- {e^{-\phi_\infty} Q^2 \over 2mr_B} \left(1 - 
{2me^{\phi_\infty} \over r_B}\right) \no\\
&&+\sqrt{\left(1-{Q^2 e^{-\phi_\infty} \over mr_B}\right) \left(1- 
{2me^{\phi_\infty} \over r_B}\right)}\Bigg] \label{M1b-string} \,.
\end{eqnarray}
In the limit $r_B \to \infty$, $\bar {M}(r_B) \to m$.

We next study the Einstein-frame solution that is conformally related to 
(\ref{4Ddilbh}) through the conformal transformation (\ref{auxmet}), where
\be
\label{CT2}
U = {e}^{-2\phi_\infty}
\left(1- {Q^2 {e}^{-\phi_\infty} \over {mr}}\right) \,.
\ee
Thus, the Einstein metric is
\begin{eqnarray}
{ds}^2 &=& -\left( 1-{2me^{\phi_\infty} \over r}\right){dt}^2 +
{e}^{-2\phi_\infty} \left(1- {2me^{\phi_\infty} \over r}\right)^{-1} {dr}^2 
\no \\
&&+ {e}^{-2\phi_\infty} r^2 \left( 1- {Q^2 e^{-\phi_\infty} \over mr}\right)
{d\omega}^2 \,. \label{GHSE}
\end{eqnarray}
Once again, since the above spacetime is asymptotically flat, we choose the
reference metric to be flat:
\be
\label{GHSE-flat}
{ds}^2_0 = -{N}^2_0 {dt}^2 + d\rho^2 + \rho^2 {d\omega}^2 \ \ ,
\ee
where $\rho$ is the radial coordinate and $N_0 =$ constant. In the 
above coordinates, a two-sphere (with $t$ and $\rho$ constant) embedded in 
this reference spacetime is not isometric with a two-sphere (with $t$ and 
$r$ constant) embedded in spacetime (\ref{GHSE}). However, they can be made
isometric by defining $\rho$ in terms of the curvature coordinate $r$ as 
\be
\label{r-rho}
{\rho} = r \left( 1- {Q^2 {e}^{-\phi_\infty} \over mr_B}\right)^{1/2} 
{e}^{-\phi_\infty} \,.
\ee
One can implement this coordinate transformation in either Eq. (\ref{GHSE}) 
or (\ref{GHSE-flat}). Both choices yield the same mass expressions. We choose
to apply it in Eq. (\ref{GHSE-flat}). In these coordinates, the flat metric 
gets recast to 
\be
\label{GHSE-ref}
{ds}^2_0 = -{N^2}_0 {dt}^2 + {e}^{-2\phi_\infty}\left(1-{Q^2 
{e}^{-\phi_\infty} \over mr_B}\right) dr^2 + r^2 \left(1-{Q^2
{e}^{-\phi_\infty} \over mr_B}\right) {e}^{-2\phi_\infty} d{\omega}^2.
\ee
For matching the lapse on the boundary at $r=r_B$, we require
\be
\label{lapse1-ref}
N_0 = \left( 1- {2m{e}^{\phi_\infty} \over r_B}\right)^{ 1/2}
\ee 
everywhere. Note that the flat metric (\ref{GHSE-ref}) is indeed conformally
related to the reference solution (\ref{GHS-ref}) in the string frame.
By the application of (\ref{BLM}), we find that the 
quasilocal mass turns out to be that given in (\ref{M1b-string}). This shows 
that the quasilocal mass at any $r$ is a conformal invariant. 

We now consider electrically charged black hole solutions of the action
(\ref{S4Ddil}). The associated metric, dilaton, and the non-vanishing
Maxwell field tensor components are 
\begin{eqnarray}
d \bar{s}^2 &=& -{e^{-\phi_\infty}\left(1+ (Q_e^2 -2m^2 
e^{2\phi_\infty}) /(me^{\phi_\infty} r)\right) 
\over \left(1+Q^2 /(m e^{\phi_\infty} r) \right)^2 }dt^2 \nonumber \\    
&& + {dr^2 \over \left(1+(Q_e^2 -2m^2 e^{2\phi_\infty}) /(me^{\phi_\infty} r)
\right)}  +r^2 d\omega^2 
\label{4Ddilbhes} \ \ , \\
U(\phi) &=& e^{-2\phi}= \left( 1 + {Q^2_e e^{-\phi_\infty} \over mr} \right) 
\label{4Ddile} \ \ , \\
\bar{F}_{tr}&=&{Q_e e^{4\phi} \over r^2 } \,.
\label{4DFes}
\end{eqnarray}
Since this spacetime is asymptotically flat, we choose the reference solution
to be flat with the metric (\ref{GHS-ref}), where the lapse is just 
$\sqrt{-\bar{g}_{tt}}$ in Eq. (\ref{4Ddilbhes}) evaluated at $r=r_B$.

By applying our prescription for finding the quasilocal mass (as we did in the 
case of the magnetically charged black holes) to this case, 
we find that in the string frame
\be\label{M1es}
\bar{M}(r_B) = {e}^{-\phi_\infty} r_B \Bigg\{\bar{\lambda} 
- \bar{\lambda}^2  + {Q_e^2 \bar{\lambda}^2 \over 2m{e}^{
\phi_\infty}r_B} \left(1+{Q_e^2 \over m{e}^{
\phi_\infty}r_B}\right)^{-1} \Bigg\} \ \ ,
\ee
where $\bar{\lambda}^{-2} \equiv \bar{g}_{rr} (r_B )$ in Eq. 
(\ref{4Ddilbhes}). Thus, the total mass of the spacetime is once again $m$.

On the other hand, in the Einstein frame the metric is 
\begin{eqnarray}
d{s}^2 &=& -e^{-\phi_\infty}\left(1-  {2m^2 
e^{2\phi_\infty} \over me^{\phi_\infty} r + Q_e^2 }\right) dt^2 \nonumber \\
&&+ \left(1-  {2m^2 e^{2\phi_\infty} \over me^{\phi_\infty} r + Q_e^2 }
\right)^{-1} dr^2
+ r^2 \left(1+ {Q_e^2 \over me^{\phi_\infty} r }\right) d\omega^2 
 \ \ , \label{4DdilbheE}
\end{eqnarray}
which is related to the string metric (\ref{4Ddilbhes}) via the conformal
transformation (\ref{auxmet}), where $U$ is given by (\ref{4Ddile}). Since the 
above solution is asymptotically flat, the reference metric is chosen to be 
flat once again:
\be
d {s}^2_0 = -{N}_0^2 dt^2 + \left(1+ {Q_e^2 \over me^{\phi_\infty}r_B 
}\right) dr^2 + r^2_B \left(1+ {Q_e^2 \over me^{\phi_\infty} r_B }\right) 
d\omega^2 \ \ ,
\ee
where, as in (\ref{GHSE-ref}), we use coordinates such that the 2-sphere
boundary at $r=r_B$ is manifestly isometric with that in the spacetime
(\ref{4DdilbheE}). Also, $\phi_0$ is defined so as to match with the solution
(\ref{4Ddile}) at the boundary 2-sphere at $r=r_B$
\be
e^{-2\phi_0} = U(\phi_0)|_{r=r_B} = 1+Q_e^2 / (me^{\phi_\infty} r_B) \,.
\ee
This is then the value of $\phi_0$ everywhere in the reference spacetime. 
Similarly the reference lapse ${N}_0$ is chosen to be $\sqrt{-g_{tt}}$
in (\ref{4DdilbheE}) evaluated at $r=r_B$. Our prescription for evaluating the
quasilocal mass then yields the same expression as found in 
(\ref{M1es}) for the string frame, thus demonstrating its conformal 
invariance.

\subsection{Asymptotically non-flat black holes}

To demonstrate that our prescription yields a conformally invariant
definition of quasilocal mass even for asymptotically non-flat solutions
we consider a particular black hole solution of Chan, Horne, and Mann
that arises from the following action \cite{CHM}:
\be
\label{CHMacE}
S = {1\over 2 \kappa} \int{d^4 x }\sqrt{-g} \> [R -2(\nabla\phi)^2 
-{e}^{-2\phi} F^2] \,.
\ee
The fields of the electrically charged black hole solution in this theory are
\begin{eqnarray}
{ds}^2 &=& -{1 \over{\gamma^4}} (r^2 - 4{\gamma^2}M){dt^2} + 
{4r^2 \over r^2 -{4\gamma^2}M }{dr^2} + r^2 d\omega^2 \label{CHMbhE} \ \ ,\\
{e}^{-2\phi} &=& {2Q^2 \over r^2} \label{CHMdil} \ \ ,\\
F_{tr} &=& {r \over 2Q\gamma^2} \label{CHMF} \ \ ,
\end{eqnarray}
where $\gamma$ is a constant with dimensions of $\sqrt{r}$ and $Q$ is the 
electric charge. 

In this case, there is no unique way to choose the reference geometry. Here,
we choose to compare the quasilocal mass of the above solution with respect to
a geometry whose (non-flat) space part of the metric is determined by setting 
$M=0$ in (\ref{CHMbhE}). Thus, our reference geometry is:
\be
\label{CHM-refE}
{ds^2}_0 = - N_0^2 dt^2 + 4{dr}^2 + r^2d{\omega}^2.
\ee
For the 2-sphere boundary at $r = r_B$, the HH prescription dictates that
\be \label{N2}
N_0 = {1 \over \gamma^2} ({r^2}_B - 4\gamma^2 M)^{1/2}
\ee
is obeyed everywhere. Our prescription for the quasilocal mass then yields
\be
\label{MCHME}
M(r_B) = {r_B^2 \over 2{\gamma}^2} \left[ 
\sqrt{ 1-{4{\gamma}^2 M \over 
{r^2}_B}} - 1 + {4{\gamma}^2 M \over {r^2}_B} \right]
\,. \ee
As $r_B \to \infty$, $M(r_B) \to M$.

We next consider the string action conformally related to (\ref{CHMacE}):
\be
\label{CHMacs}
S = {1 \over 2{\kappa}} \int d^4 x \sqrt{-\bar g} e^{-2\phi}
[\bar R + 4 (\bar{\nabla} \phi)^2 - \bar{F}^2]
\ee  
where the conformal factor is given by
\be
\label{CT-CHM}
{\Omega} = e^{\phi} = {r \over \sqrt{2} Q} \>\>\> \left( = U^{-1/2}\right) \,.
\ee
Therefore, the string metric conformally related to (\ref{CHMbhE}) is
\be
\label{CHMs}
d\bar{s}^{2} =  -{r^2 \over 2Q^2 {\gamma}^4} ( r^2 - 4{\gamma}^2 M ) dt^2 +
{2r^2/Q^2 \over 1-4{\gamma}^2 M/r^2}dr^2 + {r^4 \over 2Q^2} d{\omega}^2 \,.
\ee
The space part of the reference string metric is chosen by setting $M=0$
above, which gives the reference metric to be:
\be
\label{CHM-refs}
ds_0^2 = - \bar{N}_0^2 dt^2 +{ 2r^2 \over Q^2} dr^2 + 
{r^4 \over 2Q^2} d{\omega}^2 \ \ ,
\ee
where the reference lapse is obtained by matching with that in Eq. 
(\ref{CHMs}) at the boundary $r=r_B$.
\be
\bar N_0 = {r_B \over \sqrt{2} Q\gamma^2} ({r^2}_B - 4\gamma^2 M)^{1/2} \,.
\ee
With these prescribed choices for the reference fields, we find that the
quasilocal mass is
\be
\label{MCHMs}
\bar M (r_B) = {  r_B \over 2{\gamma}^2 } \left[ 1- \sqrt{1-4{\gamma}^2 M 
\over {r^2}_B}\right] ({r^2}_B - 4\gamma^2 M )^{1/2}  \ \ ,
\ee
which is the same as that evaluated in the Einstein frame, namely,
Eq. (\ref{MCHME}).

\section{Discussion}
\label{sec:disc}

Naive expectations from quantum field theory would suggest that physical 
quantities should remain invariant under a conformal transformation. However, 
when it comes to the behavior of quasilocal mass under such a transformation, 
one must bear caution. This is because {\em a priori} it can not be ruled out
that in some frames the scalar field $\phi$, which defines the conformal 
factor, itself contributes to the energy-momentum of the spacetime. In this 
paper we showed that, the preceding caveat notwithstanding, the unreferenced 
BY quasilocal mass is indeed conformally invariant.

However, to obtain the physical mass of a spacetime one is often required to 
subtract a reference term. At the level of the action, this is achieved by 
subtracting  a reference action. Different choices of reference action will
lead to different physical masses for the same classical solution. Moreover,
the reference term $\bar{M}_0$ arising from such actions and, consequently, 
the referenced quasilocal mass may not be conformally invariant.   

In this paper, we attempted to reduce the arbitrariness in the choice of a 
reference action. We motivated this choice from a basic principle, in the form
of the Hawking-Horowitz prescription, which requires the reference geometry to 
obey certain conditions. We proved that this prescription automatically gives 
rise to a conformally invariant referenced quasilocal mass if the conformal 
factor is monotonic in the scalar field.

We note, however, that the HH prescription does not attempt to specify a 
unique reference geometry, owing to which the referenced quasilocal mass 
is non-unique, albeit conformally invariant. It is only in some special 
cases that one can obtain a unique physical mass. Asymptotically flat 
spacetime solutions of general relativity belong to this category. There the 
positive energy theorem and the stability criterion for Minkowski spacetime 
ensure that under certain positivity conditions on the energy-momentum 
tensor, the total energy of such spacetimes is positive; it is zero only for 
the Minkowski spacetime. This selects the flat spacetime as a very 
special reference geometry for calculating the total energy and, in certain 
cases, the quasilocal mass/ energy of regions in such spacetimes. The 
conformal invariance of quasilocal mass implies that in conformally related 
spacetimes, which are asymptotically flat, the flat spacetime continues 
to be a special reference geometry.

In this vein, one may argue that if the positive energy theorem could be 
shown to hold for asymptotically non-flat cases, atleast of a limited type 
such as the SSS spacetimes, then a corresponding special reference geometry 
may emerge, which could be used under the HH prescription to compute the 
referenced quasilocal mass in such spacetimes in some unique way. This and 
other related issues are currently under study \cite{SB}.

\section{Acknowledgments}

We thank Valerio Faraoni, Sayan Kar, Jorma Louko, and Sukanya Sinha for 
helpful discussions. One of us (S. B.) is grateful to the members of the 
Theoretical Physics Group at the Raman Research Institute, Bangalore, India, 
where part of this work was done, for their hospitality. We would also like to
thank IUCAA for financial support.

\appendix\section{The stress-energy pseudotensor}
\label{app:PT}

In this section we present an alternative derivation of the quasilocal energy
expression (\ref{energyBD}) using the pseudotensor method \cite{LL,ABS,MTW}.
The outline of our approach, as applied to the scalar-tensor type theories, 
(\ref{SBD}), is as follows. As is the case for the Brans-Dicke theory, even 
in our generalized non-minimally coupled theory, we make use of the Bianchi
identity and the equations of motion to show that the covariant divergence
of the matter stress-tensor vanishes. In a particular coordinate system,
this is then shown to imply that the ordinary divergence of the sum
of two quantities vanishes. One of the terms in this sum is the matter
stress-tensor in that coordinate frame, and the other term is then
interpreted as the stress-energy contribution of the geometry (that is,
of the gravitational as well as the Brans-Dicke type scalar field). 
Naturally, any statement made about the stress-energy content of the
geometry using this approach will be frame-dependent. However, for 
spherically symmetric cases, a meaningful proper energy contained in the 
sphere $S^D$ with its origin at the center of spherical symmetry,
can be defined using this method \cite{MTW}. It is in this context that we 
present this alternative derivation of the quasilocal energy expression 
Eq. (\ref{energyBD}).

We begin by finding the equations of motion for the theory described by 
(\ref{SBD}) for the special case where $W(\phi) = X(\phi)$ and $U(\phi)W(\phi) 
= 1$. (Our results can be generalized in a straightforward 
manner to other cases not constrained by these conditions.) Requiring 
the action to be stationary under variations of the metric tensor and the 
field $\phi$, gives the equations of motion:
\be
\label{eom1}
U(\phi)[\bar{R}^{\mu\nu} - {1\over 2}g^{\mu\nu} \bar{R}] = -{1\over 2}
[\bar{T}_{\rm m}^{\mu\nu} + T_\phi^{\mu\nu} + 2U(\phi)^{;\mu;\nu} 
- 2\bar{g}^{\mu\nu}U(\phi)]^{;\lambda}_{;\lambda}] \ \ ,
\ee
\be
\label{eom2}
\bar{g}^{\mu\nu}\phi_{;\mu;\nu} -2 {\delta (UV) \over {\delta\phi}} 
- \bar{R}{\delta U \over {\delta \phi}} = 0 \ \ ,
\ee
where we have used the notation $A_{,\mu} \equiv \partial A / \partial 
x^{\mu}$ and $A_{;\mu} \equiv \bar{\nabla}_\mu A$. Here, 
$\bar{T}_{\rm m}^{\mu\nu}$ is the energy-momentum tensor of matter obtained
by varying $\bar{L}_{\rm m}$ with respect to $\bar{g}_{\mu\nu}$, and
\be
\label{Tphi}
T_\phi^{\mu\nu} \equiv -2 \partial^\mu\phi\partial^\nu\phi - 
\bar{g}^{\mu\nu}
[-\partial^\lambda\phi\partial_\lambda\phi -U(\phi)V(\phi)]
\,.
\ee
We have considered, for the present, a $\phi$-independent $\bar{L}_{\rm m}$. 
Since our discussion here is limited to the ``Brans-Dicke'' frame only, 
we shall henceforth drop the overbar.

It would be 
reasonable to demand that this theory conforms to the equivalence principle. 
To demonstrate that this indeed holds i.e. $T^{\mu\nu}_{{\rm m}\>;\nu} = 0$, 
we first note that the contracted 
Bianchi identity satisfied by the Einstein tensor gives:
\be
\label{Bianchi}
\left[  U(\phi)^{-1} (T_{\rm m}^{\mu\nu} + t^{\mu\nu} )\right]_{;\nu} = 0
\ee
with 
\be
\label{t}
t^{\mu\nu} \equiv T_\phi^{\mu\nu} +  2U(\phi)^{;\mu;\nu} -
2g^{\mu\nu}U(\phi)^{;\lambda}_{;\lambda} \,
\ee
{}From the equation of motion (\ref{eom1}) and the Bianchi identity it 
follows that
\be
\label{coveom1}
U(\phi)_{,\nu}[R^{\mu\nu} - {1\over 2}g^{\mu\nu}R] = 
-{1\over 2}[T^{\mu\nu}_{{\rm m}\>;\nu} + t^{\mu\nu}_{;\nu}] \,.
\ee
The definition of the Riemann tensor yields the identity 
\be
\label{Rid}
R_{\rho\alpha} U(\phi)^{;\rho} = U(\phi)^{;\lambda}_{;\lambda ;\alpha}
- U(\phi)^{;\lambda}_{;\alpha;\lambda}
\,.
\ee
Taking the covariant derivative of (\ref{Tphi}) and using the equation of 
motion (\ref{eom2}) for the scalar field $\phi$, we obtain
\be
\label{divTp}
T^{\mu\nu}_{\phi\>;\nu} = U(\phi)^{;\mu} R  
\,.
\ee
Taking the covariant derivative of (\ref{t}) and using Eqs. (\ref{Rid}) and 
(\ref{divTp}) then gives the desirable vanishing covariant divergence of the
matter stress energy tensor.

    Our task now is to find the expression for a conserved stress-energy 
pseudotensor for the geometry. To achieve this we proceed by expressing 
the vanishing of the
covariant divergence of the matter stress-energy tensor as:
\be
\label{ordBianchi}
[\sqrt{-g}T^\nu_{{\rm m}\>\mu }]_{,\nu} 
- {1\over 2}g_{\tau\beta,\mu} \sqrt{-g}T^{\tau\beta}_{\rm m} = 0 
\ \ .
\ee
To cast the left-hand side of the above equation into a total ordinary 
divergence one has to seek a representation of the second quantity in terms of
a total ordinary divergence. This can be done as follows. First, we make use 
of the equation of motion (\ref{coveom1}) to express the matter stress-energy 
tensor in terms of purely geometrical quantities, namely, the scalar field
and metric-dependent quantities:
\be
\label{Tm}
T_{\rm m}^{\tau\beta} = -t^{\tau\beta} -2U(\phi)G^{\tau\beta} \ \ ,
\ee
where $G^{\tau\beta}$ is the Einstein tensor. (For simplicity, we have above 
dropped a possible contribution from a divergenceless term.)
Second, note that the right-hand side of this expression is merely the 
functional derivative of 
\be
J \equiv 2\int \>d^{(D+1)} x\>\sqrt{-g}\>[U(\phi )R + L_\phi ]
\ee
under variations of the metric tensor, with boundary conditions that require 
the vanishing of the metric and its first derivatives on the boundary of a 
$(D+1)$-dimensional manifold over which this  integral has been taken. For 
more general variations, one would get contributions from the surface 
integrals as well. Here, the variation of $L_{\phi }$ with respect to
the metric tensor is taken to yield $T_{\phi }^{\mu\nu}$. We consider the 
standard decomposition of $\sqrt{-g}R$ into a pure divergence term and a 
simple expression involving only the metric and its first derivatives:
\be
\sqrt{-g}R = \Alpha + 
[\sqrt{-g}g^{\sigma\rho}\Gamma^\alpha_{\sigma\alpha}]_{,\rho}
-  [\sqrt{-g}g^{\sigma\rho}\Gamma^\alpha_{\sigma\rho}]_{,\alpha}
\ee
with 
\be
\label{calU}
\Alpha \equiv
\sqrt{-g}g^{\sigma\rho}[\Gamma^\alpha_{\sigma\rho}\Gamma^\beta_{\alpha\beta}
- \Gamma^\alpha_{\beta\rho}\Gamma^\beta_{\alpha\sigma}] \,.
\ee
It follows that the functional derivative of $J$ with respect to the 
metric tensor is the same as that of  
\be
H \equiv \int d^{(D+1)} \>x\> [\Beta + \sqrt{-g}L_\phi] 
\ \ ,
\ee
where
\be
\label{hatV}
{\cal V} \equiv [U\Alpha 
- \sqrt{-g}g^{\sigma\rho}\Gamma^\alpha_{\sigma\alpha}U_{,\rho}
+ \sqrt{-g}g^{\sigma\rho}\Gamma^\alpha_{\sigma\rho}U_{,\alpha}] \,.
\ee
Comparing the variations of $J$ and $H$ with respect to the metric, we get:
\begin{eqnarray}
\sqrt{-g}UG_{\mu\nu} + \sqrt{-g}[U_{;\mu;\nu} - 
g_{\mu\nu}U^{;\alpha}_{;\alpha}]
&+& {1\over 2}\sqrt{-g}T_{\phi \>\mu\nu} \nonumber \\
&=& {\partial (\Beta + \sqrt{-g}L_\phi)\over \partial g^{\mu\nu}}
- \left[{\partial(\Beta + \sqrt{-g}L_\phi ) \over 
{\partial g^{\mu\nu}_{,\lambda}}} \right]_{,\lambda} \ \ ,
\end{eqnarray}
where we made use of Eqs. (\ref{Tm}) and (\ref{t}). Next, we define 
\be
\label{hatVdef}
\hat \Beta \equiv \Beta + \sqrt{-g}L_\phi \,. 
\ee
The expression for the ordinary derivative of $\hat \Beta$ and the field 
equation (\ref{eom2}) for $\phi$ enable us to express 
$g_{\tau\beta ,\mu} [t^{\tau\beta} +2U(\phi)G^{\tau\beta}]$ as a total 
divergence. Using Eq. (\ref{Tm}), we express Eq. (\ref{ordBianchi}) as
a vanishing total derivative:
\be
\left[\sqrt{-g}T^\nu_{{\rm m} \>\mu} - \hat \Beta \delta^\nu_\mu
- {\partial{\hat\Beta}\over {\partial g^{\tau\beta}_{,\nu}}} \>
g^{\tau\beta}_{,\mu} -{{\partial \hat\Beta}\over {\partial\phi_{,\nu}}}
\> \phi_{,\mu} \right]_{,\nu} = 0 \,.
\ee
For $\nu =0$, the expression within the brackets integrated over
a spacelike hypersurface is thus invariant under time translations for 
a distribution of matter with a compact support over the surface. 
This is the expression for the stress-energy pseudotensor that we seek.
The quantity
\be
\label{em4vector}
P_\mu \equiv {1\over 2\kappa} \int_{\Sigma} d \Sigma \left[\sqrt{-g} \>
T^0_{{\rm m} \>\mu} - \hat \Beta \delta^0_\mu
- {\partial{\hat\Beta}\over {\partial g^{\tau\beta}_{,0}}} 
\> g^{\tau\beta}_{,\mu}
-{{\partial \hat\Beta}\over {\partial\phi_{,0}}} \> \phi_{,\mu} \right]
\ \ ,
\ee
evaluated on a constant-time spacelike hypersurface $\Sigma$, is thus 
conserved. 
This may be viewed as the generalization of the energy momentum four vector
for a Brans-Dicke theory. As in general relativity, $P_\mu$
is not a generally covariant four-vector since $\Alpha$ and $\Beta$ are not 
scalar densities. The intrinsic non-covariance of the energy-momentum 
density of the gravitational field has its origin in the intimate 
connection between geometry and the gravitational field. Had the expression
been covariant, one could always have gone into a preferred (freely falling) 
frame to ensure vanishing of an arbitrary localized gravitational field.

       The above form for the energy-momentum pseudotensor for the 
generalized Brans-Dicke theory can also be obtained by considering a variation
of the coordinate system instead of the metric field. The analysis enables us 
to express the gravitational stress-energy pseudotensor in a very compact 
form, which is identical to the expression derived in the quasilocal 
formalism. To demonstrate this, we consider 
\be
\label{HV}
H = \int \hat {\cal V} d^{D+1)}x \ \ ,
\ee
where $\hat{\cal V}$ is a function of the metric, the scalar field $\phi$, 
and their first derivatives. Its variation is:
\be
\label{dV}
\delta\hat{\cal V} = {\partial\hat{\cal V}\over \partial g^{\mu\nu}}\delta 
g^{\mu\nu} + {\partial\hat{\cal V}\over \partial g^{\mu\nu}_\lambda}\delta 
g^{\mu\nu}_\lambda + {\partial\hat {\cal V}\over \partial\phi}\delta\phi
+ {\partial\hat {\cal V}\over \partial\phi_{,\lambda}}\delta\phi_{,\lambda}
\,.
\ee

Consider an infinitesimal change of coordinates of the form:
\be
\label{coordt}
\hat x^\alpha = x^\alpha + \epsilon\xi^\alpha \,.
\ee
Retaining terms up to the first order in $\epsilon$, we get the following 
variations:
\be
\label{xct}
{\partial x^\alpha\over \partial\hat x^\lambda} = \delta^\alpha_\lambda 
- \epsilon {\partial \xi^\alpha\over \partial x^\lambda} + O(\epsilon^2)
\ \ ,
\ee
\be
\label{gct}
\delta g^{\mu\nu} = \epsilon(\xi^\mu_{,\alpha}g^{\alpha\nu} +\xi^\nu_{,\alpha}
g^{\alpha\mu}) \ \ ,
\ee
\be
\label{dgct}
\delta g^{\mu\nu}_{,\lambda} = \epsilon \left( g^{\tau\nu}_{,\lambda}
\xi^\mu_{,\tau} + g^{\mu\beta}_{,\lambda}\xi^\nu_{,\beta}
- g^{\mu\nu}_{,\alpha}\xi^\alpha_{,\lambda}
+ g^{\tau\nu}\xi^\mu_{,\tau ,\lambda}
+ g^{\tau\mu}\xi^\nu_{,\tau ,\lambda} \right) \ \ ,
\ee
\be
\label{sqgct}
\delta\sqrt{-g} = -\epsilon\sqrt{-g}\xi^\alpha_\alpha \ \ ,
\ee
\be
\label{phict}
\delta\phi = 0 \ \ ,
\ee
\be
\label{dphict}
\delta(\phi_{,\lambda}) = -\epsilon\phi_{,\alpha}\xi^\alpha_{,\lambda} \ \ ,
\ee
A restriction to linear transformations enables one to get an elegant form for
$\delta\hat {\cal V}$. The Christoffel symbols transform as tensors under such 
transformations and hence $\hat {\cal V}$ transforms as a scalar density. 
Thus 
\be
\label{Vct}
\delta\hat {\cal V} = {\hat {\cal V}\over \sqrt{-g}}\delta\sqrt{-g} = 
-\epsilon\xi^\alpha_{,\alpha}\hat {\cal V} \,.
\ee
Substituting the variations (\ref{gct})-(\ref{dphict}) for an arbitrary linear 
coordinate transformation into (\ref{dV}), and comparing the expression with 
(\ref{Vct}), we obtain the identity:
\be
\label{Vid}
{\partial\hat{\cal V}\over \partial g^{\mu\nu}}g^{\alpha\nu}
+ {\partial\hat{\cal V}\over \partial g^{\mu\nu}_{,\lambda}}
g^{\alpha\nu}_{,\lambda} -{1\over 2} {\partial\hat{\cal V}\over \partial 
g^{\beta\nu}_{,\alpha}}g^{\beta\nu}_{,\mu}
- {\partial\hat {\cal V}\over \partial\phi_{,\alpha}}\phi_{,\mu}
= - {1\over 2}\hat {\cal V} g^\alpha_\mu \,.
\ee
Although the above identity was derived for variations under linear coordinate
transformations, one can verify that it holds quite generally \cite{ABS}.
The use of this identity yields a simple expression for the variation of 
$\hat {\cal V}$ under the general transformation (\ref{coordt}):
\be
\label{dVid}
\delta\hat {\cal V} = -\epsilon\hat {\cal V}\xi^\alpha_{,\alpha} + 
2\epsilon{\partial\hat {\cal V}\over \partial g^{\mu\nu}_{,\lambda}}
\xi^\mu_{,\tau ,\lambda}g^{\tau\nu} \,.
\ee

Under conditions where $\xi$ and its derivatives are taken to vanish on the 
boundary, the variation of the metric tensor and its derivatives also vanish 
there. Under such boundary conditions, H has a vanishing variation, i.e.,
\be
\label{dH}
\delta H = \int_\Sigma\delta \left({\hat {\cal V} \over \sqrt{-g}} \right)
\sqrt{-g}\> d^{(D+1)}x = 0 \ \ ,
\ee
which, using the above identities, reduces to
\be
\label{dHid}
\delta H = 
2\epsilon\int_\Sigma {\partial\hat {\cal V}\over \partial 
g^{\mu\nu}_{,\lambda}}\xi^\mu_{,\tau ,\lambda}g^{\tau\nu}d^{(D+1)}x = 0 \,.
\ee
This expression may be integrated by parts twice. Since $\delta H$ vanishes 
for arbitrary $\xi^\mu$, we obtain the following divergence law:
\be
\label{div}
\left({\partial\hat {\cal V}\over \partial g^{\mu\nu}_{,\lambda}}
g^{\tau\nu}\right)_{,\tau ,\lambda} = 0 \,.
\ee
Thus 
\be
\label{F}
\sqrt{-g}F^\tau_\mu \equiv \left({\partial\hat {\cal V}\over \partial 
g^{\mu\nu}_{,\lambda}} g^{\tau\nu} \right)_{,\lambda} 
\ee
defines a conserved quantity. Using the identity (\ref{Vid}) and the field 
equation (\ref{eom1}) gives:
\be
\label{Fpseudo}
\sqrt{-g}F^\tau_\mu = -\sqrt{-g}T^{(m)\tau}_\mu -{1\over 2}\hat {\cal V} 
g^\tau_\mu + {1\over 2}({\partial\hat {\cal V}\over \partial 
g^{\beta\nu}_{,\tau}}g^{\beta\nu})_{,\mu} -{1\over 2}{\partial\hat {\cal V}
\over \partial\phi_{,\tau}}\phi_{,\mu} \ \ ,
\ee
which is just the expression that we had obtained for the stress energy 
pseudotensor by the variation of the metric tensor earlier. The expression 
(\ref{div}) for a vanishing ordinary divergence implies that 
\be
\label{PV}
P_\mu \equiv - {1\over \kappa} \int_V \left({\partial\hat {\cal V}\over 
  \partial g^{\mu\nu}_{,\lambda}}g^{o\nu} \right)_{,\lambda}dV 
\ee
is a conserved quantity if $V$ is the entire space at a given time. In the 
special case of a time independent metric, Gauss's theorem in $D$ dimensions 
gives the energy momentum as surface integral over a $(D-1)$-dimensional 
surface:
\be
\label{PS}
P_\mu = -{1\over \kappa} \int_\Sigma({\partial\hat {\cal V}\over \partial 
     g^{\mu\nu}_{,j}} g^{o\nu})d\Sigma_j \,.
\ee
This gives the interesting result that in the generalized Brans-Dicke theory, 
the generalized energy-momentum in a $D$-dimensional volume can be determined 
by the metric-tensor and its derivatives on the $(D-1)$-dimensional 
surface, the details of the field inside the volume being irrelevant.

We conclude this section by noting that in the curvature coordinates, the 
above expression (\ref{PS}) evaluated for the SSS metric (\ref{SSS}) tallies 
with the quasilocal energy expression (\ref{energyBD}) (with $\bar{E}_0$ set
equal to zero). This can be seen from the definition of $\cal V$ given by Eqs.
(\ref{hatVdef}) and (\ref{hatV}): The term $U\bar{k}$ in (\ref{energyBD}) 
yields the first term in the definition
(\ref{hatV}) of ${\cal V}$, whereas the term $\bar{n}^i \partial_i U$ in 
(\ref{energyBD}) yields the second and the third terms of ${\cal V}$.

\vfil
\pagebreak

\centerline{FIGURE CAPTION}
\vskip 0.2in

Figure 1: A bounded spacetime region with boundary consisting of initial
and final spatial hypersurfaces $t=t_1$ and $t=t_2$ and a  
$D$-dimensional surface ${}^D {\sf B}$. Here, ${}^D {\sf B}$ itself is 
the time-evolution
of the $(D-1)$-dimensional surface ${\sf B}$, which is the boundary of an 
arbitrary spatial slice $\Sigma$.
\vfil
\pagebreak

\begin{figure}[tb]
\begin{center}\leavevmode\epsfbox{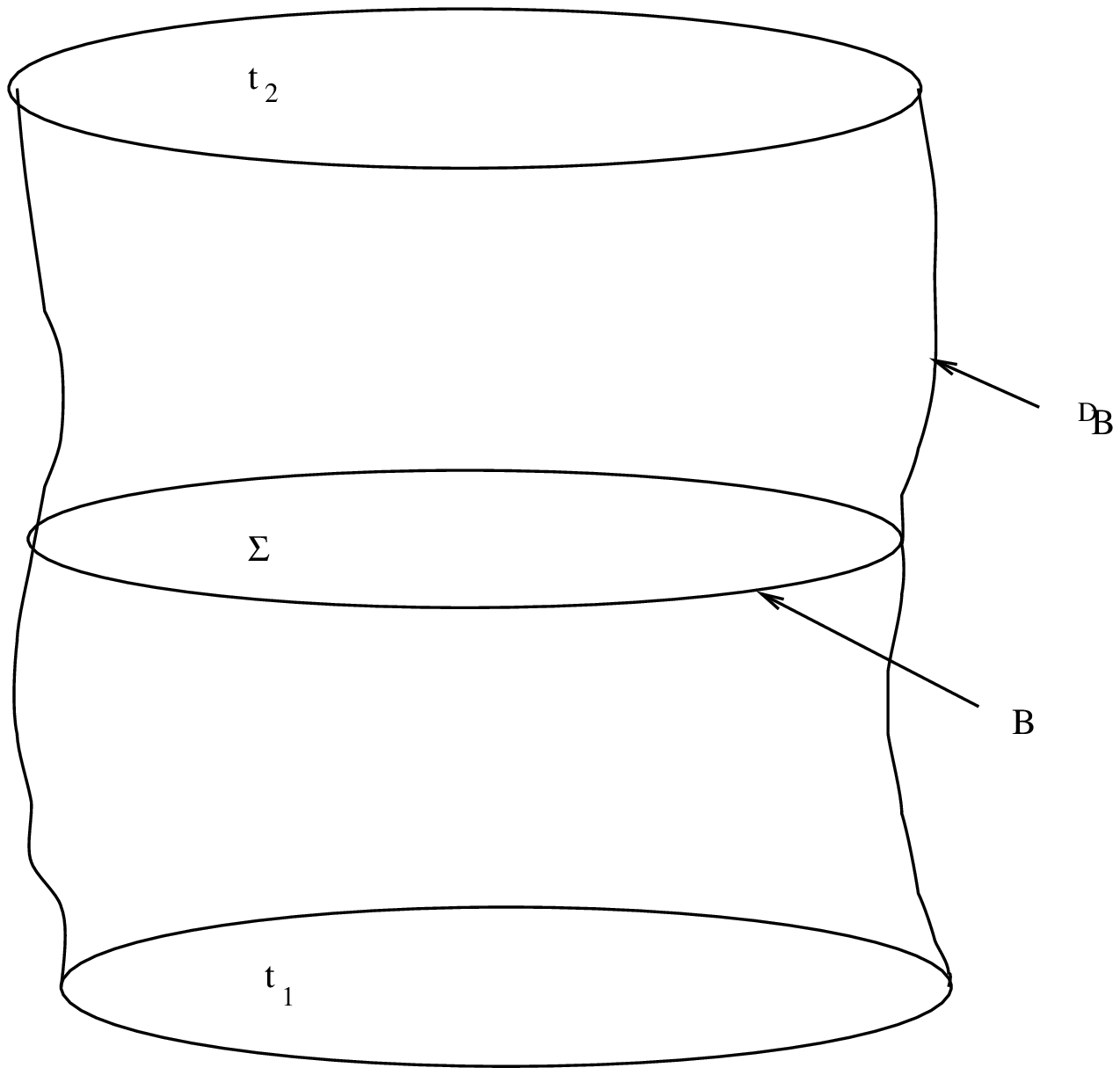}\end{center}
\caption{}
\label{fig1}
\end{figure}

\end{document}